\documentclass[twocolumn]{aastex61}
\hypersetup{linkcolor=red,citecolor=cyan,filecolor=cyan,urlcolor=magenta}

\newcommand{\water}{H$_2$O}
\newcommand{\methane}{CH$_4$}

\shorttitle{Exo-Neptune Demographics}
\shortauthors{Crossfield \& Kreidberg}

\begin{document}


\title{Trends in Atmospheric Properties of Neptune-Size Exoplanets}


%% \author{
%% Ian J.\ M.\ Crossfield\altaffilmark{1,2}, 
%% Laura Kreidberg\altaffilmark{3,4}}
%% %% Notice that each of these authors has alternate affiliations, which
%% %% are identified by the \altaffilmark after each name.  Specify alternate
%% %% affiliation information with \altaffiltext, with one command per each
%% %% affiliation.
%% \altaffiltext{1}{Department of Physics, Massachusetts Institute of Technology,
%% Cambridge, MA, USA; \url{iancross@mit.edu} }
%% \altaffiltext{2}{Astronomy and Astrophysics Department, UC Santa Cruz, CA, USA}
%% \altaffiltext{3}{Center for Astrophysics,
%%     60 Garden Street, Cambridge, MA, USA; \url{laura.kreidberg@cfa.harvard.edu}}
%% \altaffiltext{4}{Society of Fellows, Harvard University.}

\author{Ian J.\ M.\ Crossfield}
\affiliation{Department of Physics, Massachusetts Institute of Technology,
  Cambridge, MA, USA; \href{mailto:iancross@mit.edu}{iancross@mit.edu}}
\affiliation{Astronomy and Astrophysics Department, UC Santa Cruz, CA, USA}

\author{Laura Kreidberg}
\affiliation{Center for Astrophysics, 60 Garden Street, Cambridge, MA, USA; \href{mailto:laura.kreidberg@cfa.harvard.edu}{laura.kreidberg@cfa.harvard.edu}}
\affiliation{Society of Fellows, Harvard University.}


\begin{abstract}
%Exoplanets with radii 2--6 times the size of Earth are an abundant outcome of planet
%formation, occurring around more than a quarter of all stars. 
%Formation models predict a diversity of atmospheric compositions for these planets, and an increase in metallicity toward lower planet masses. 
Precise atmospheric observations have been made for a growing sample
of warm Neptunes.  Here we investigate the correlations between these
observations and a large number of system parameters to show that, at
95\% confidence, the amplitude of a warm Neptune's spectral features
in transmission correlates with either its equilibrium temperature
($T_{eq}$) or its bulk H/He mass fraction ($f_{HHe}$) --- in addition to the standard $kT/\mu g$ scaling. These
correlations could indicate either more optically-thick,
photochemically-produced hazes at lower $T_{eq}$ and/or
higher-metallicity atmospheres for planets with smaller radii and
lower $f_{HHe}$.
%Since hazes must exist in some of these planets, we favor the former explanation.  
We derive an analytic relation to estimate the observing time needed
with JWST/NIRISS to confidently distinguish a nominal gas giant's
transmission spectrum from a flat line. Using this tool, we show that
these possible atmospheric trends could reduce the number of expected
TESS planets accessible to JWST spectroscopy by up to a factor of
eight. Additional observations of a larger sample of planets are
required to confirm these trends in atmospheric properties as a
function of planet or system quantities. If these trends can be
confidently identified, the community will be well-positioned to
prioritize new targets for atmospheric study and eventually break the
complex degeneracies between atmospheric chemistry, composition, and
cloud properties.
\end{abstract}

\keywords{planets and satellites: gaseous planets --- planets and satellites: atmospheres --- eclipses --- methods: statistical }

\vspace{-1in}
\section{Introduction}

%"We need to find things that scale with other things" (R.\ Murray-Clay).

Short-period planets with sizes of $2-6\, R_\oplus$ (hereafter, ``warm
Neptunes'') are a ubiquitous outcome of planet formation. They occur
around $>25\%$ of all stars and comprise a distinct, gas-rich
population separate from smaller, terrestrial super-Earths
\citep[e.g.,][]{buchhave:2014,fulton:2017}. Understanding this
population is therefore critical for building a comprehensive theory
of planet formation and linking the larger gas giants to the smaller
terrestrial planets. The existence of the intermediate-sized planets
raises many questions: what stunts their growth and prevents them from
reaching Jupiter proportions \citep{pollack:1996, lambrechts:2014,
  lee:2016}, or alternatively what whittles down their younger bulk to
the smaller bodies seen today \citep{owen:2013,jin:2017}?  Where did
they form in their solar systems?  Why does our solar system lack
planets in this size range?  One powerful approach to answering these
questions is to determine these planets' bulk composition --- their
core masses and the metallicity and chemistry of their outer
envelopes. These properties provide a record of the planets' origins
that can be compared to formation models.

Planet formation models predict two broad trends in atmospheric composition for warm Neptunes. One is compositional diversity, ranging from \water-rich ``super-Ganymedes'' to puffy H/He envelopes  \citep[][]{elkins-tanton:2008,fortney:2013}. Hints of this diversity appear in the mass-radius diagram for Neptune-mass planets, which shows a factor of three scatter in density \citep[][]{weiss:2014}.  Warm Neptunes likely have some hydrogen in their atmospheres \citep[][]{wolfgang:2015, rogers:2015planet}; however, there is a strong degeneracy between core mass and envelope metallicity (M/H) that prevents an exact determination of warm Neptunes' bulk makeup from mass and radius measurements alone \citep{figueira:2009,miller-ricci:2010, rogers:2011}.

%{\bf ORPHAN fragment: } Neptune-sized and smaller planets also often have other planetary companions \citep[unlike hot Jupiters;][]{huang:2016}, indicating a distinct formation pathway which further observations could reveal.

%Warm Neptunes in particular could  form either outside the ice line (as hypothesized in our Solar system) or {\em in situ} as denser, higher-M/H super-Earths with a then H/He veneer \citep{hansen:2012,dawson:2016,lee:2016}. 

 Another qualitative prediction is that the atmospheres of smaller planets should be more enhanced in metals than Jupiter-size planets \citep[e.g.][]{fortney:2013, venturini:2016}. Infalling planetesimals can ablate and pollute the atmosphere \citep[e.g][]{pinhas:2016, mordasini:2016}, so all else being equal, metal enrichment of the envelope will be more pronounced for lower-mass planets,  because they have less gas to dilute. Indeed, the metal enrichment of the Solar System gaseous planets \citep[e.g.,][]{karkoschka:2011,luszcz-cook:2013,guillot:2014} reveals a striking trend of increasing M/H with decreasing planet mass that extends to massive hot Jupiters \citep{kreidberg:2014}. Observations of lower-mass exoplanets are broadly consistent with these trends, but the uncertainties are much larger \citep{moses:2013,fraine:2014,morley:2017,wakeford:2017}. 

There has been extensive observational study of the handful of warm Neptunes that are accessible targets for atmospheric characterization with current facilities, but so far the results defy easy explanation. Some planets exhibit spectral features from water \citep{fraine:2014, wakeford:2017}, whereas others have flat, featureless spectra \citep[e.g.][]{kreidberg:2014, knutson:2014a}. In most cases, the featureless spectra could either be caused by a high mean molecular weight atmospheric composition, or high altitude clouds or hazes. Even in cases where features are detected, the features are lower in amplitude than expected for a cloud-free solar composition atmospheres \citep{fraine:2014,wakeford:2017,fu:2017}.

In this paper, we explore possible explanations for the ensemble of
warm Neptune observations. We hope to identify trends in these
planets' atmospheric properties to provide forward guidance for future
studies, especially in light of the imminent detection and
characterization efforts from the {\em TESS}, {\em CHEOPS}, and {\em
  JWST} missions \citep{broeg:2013, ricker:2014}.  We note that
several previous studies have also investigated trends in exoplanet
transmission spectra \citep{stevenson:2016b,heng:2016,fu:2017}; however, these
efforts focused mainly on hot Jupiters, whose atmospheres may differ
substantially from those of warm Neptunes.  We also note that the
sample size of warm Neptunes is small, and that previous attempts to
classify exoplanet atmospheres have not always held up as the sample
size increased or as measurements improved
\cite[e.g.,][]{hansen:2007,fortney:2008,knutson:2010,madhusudhan:2011b}. Nevertheless,
we aim to present a useful framework for discussing atmospheric trends
as the sample size grows. In Sec.~\ref{sec:obs} we present the sample
and observational data used.  We then describe our analysis of these
atmospheric measurements in Sec.~\ref{sec:analysis}
and~\ref{sec:results}, discuss the implications for the {\em
  TESS}+{\em JWST} sample in Sec.~\ref{sec:tess}, and we conclude in
Sec.~\ref{sec:conclusion}.

%The optimal choice of targets and observing strategy depends on the physical processes the shape these planets' atmospheres \citep[e.g.][]{morley:2015}.

%NASA's {\em TESS} mission is soon expected to find thousands of new, small planets around bright stars \citep{ricker:2014,sullivan:2015}. Many of these will be ideally suited for atmospheric study, but target prioritization will be challenging until we understand the processes that shape these planets' atmospheres. 
 
 %Here we examine the ensemble of observations of warm Neptunes. 

\section{Planets and Observations}
\label{sec:obs}

For the purposes of this paper, we restrict ourselves to planets with sizes $2 < R_P/R_\oplus < 6$, which are distinct from the smaller, presumably rock-dominated super-Earths \citep{fulton:2017} and yet have bulk H/He mass fractions of $<50\%$ \citep{lopez:2014}.  We also restrict our analysis to planets with $T_{eq} < 2000$~K, since small planets at these high temperatures can be significantly sculpted by atmospheric mass loss \citep{owen:2013}.   Note that both these criteria exclude 55~Cnc~e, whose bulk makeup may be consistent with little or no volatile elements \citep{demory:2016}; even if this planet has an atmosphere \citep{ridden-harper:2016,tsiaras:2016} it is highly irradiated and likely of a fundamentally different character than those in our final sample. 

Several warm Neptunes satisfying our criteria have been observed from a variety of facilities, including both broadband photometry and spectroscopy, at low and high spectral resolution, from the ground and from space, at wavelengths from the UV to the mid-IR. This is an extremely heterogeneous data set.  One concern with trying to combine such observations is that  stellar variability can introduce an arbitrary offset between observations taken at different epochs \citep[e.g.,][]{knutson:2011,fraine:2014} and a host star's non-uniform surface brightness can introduce significant slopes, especially at shorter wavelengths \citep{mccullough:2014,oshagh:2014}.  Different assumptions about fitting system parameters and instrument systematics can also introduce bias in absolute transit depth measurements \citep{stevenson:2014b}.

For these reasons we further restrict our analysis to those planets observed with a single spectroscopic instrument, {\em HST}'s Wide Field Camera 3 (WFC3), and a single grism (the near-infrared G141).
Our final sample of six planets is described in Table~\ref{tab:targets}, and spans radii from 2--6~$R_\oplus$ and temperatures from 500--1000~K.

%[describe how this approach is similar to that of Stevenson and Sing, and not too far off from that of Heng]

\begin{deluxetable*}{l r r r r r r}
\tablecaption{  Warm Neptune sample  \label{tab:targets}}
\tablehead{
   \colhead{Name}  & \colhead{$R_P$\tablenotemark{a}} & \colhead{$M_P$\tablenotemark{a}} & \colhead{$T_{eq}$\tablenotemark{b}} & \colhead{$f_{HHe}$\tablenotemark{c} }  & \colhead{\water\ amplitude}  & \colhead{WFC3 data references}\\
         & \colhead{[$R_\oplus$]} & \colhead{[$M_\oplus$]} & \colhead{[K]} & \colhead{[\%]} & \colhead{[$H_{HHe}$]} &   }
\startdata
 HAT-P-26 b & 6.15 & 18.12 & 930 & $31.7^{6.2}_{6.0}$  &  1.79 $\pm$ 0.21  & \cite{wakeford:2017}\\
 HAT-P-11 b & 4.73 & 26.19 & 810 & $15.1^{+1.8}_{2.6}$ &  1.99 $\pm$ 0.37   & \cite{fraine:2014} \\
 HD 97658 b & 2.25 &  7.56 & 690 & $1.0^{+1.0}_{1.8}$  &  -0.086 $\pm$ 0.551     & \cite{knutson:2014b} \\
   GJ 436 b & 4.22 & 23.49 & 650 & $12.0^{+1.2}_{2.1}$ &  0.46 $\pm$ 0.25   & \cite{knutson:2014a} \\
  GJ 3470 b & 4.17 & 12.90 & 620 & $12.8^{+5.2}_{5.0}$ &  0.56 $\pm$ 0.13   & \cite[][Benneke et al., submitted]{tsiaras:2017} \\
  GJ 1214 b & 2.65 &  6.45 & 530 & $3.8^{+1.3}_{-7.1}$ &  0.073 $\pm$ 0.046 & \cite{kreidberg:2014} \\
  \enddata
\tablenotetext{a}{From exoplanets.org \citep{wright:2011}.}
\tablenotetext{b}{Assuming full heat redistribution and a Bond albedo of 0.2.}
\tablenotetext{c}{From \cite{lopez:2014}.}
%\tablecaption{Warm Neptune sample and observations}
\end{deluxetable*}

\section{Analysis}
\label{sec:analysis}
Ultimately, we hope that atmospheric observations of exoplanets will provide useful measurements of  elemental and molecular abundances, atmospheric metal enrichment and chemistry,  cloud composition and particle size distribution. Beyond that, we hope to elucidate underlying trends in the ensemble properties of planetary atmospheres and learn how these are influenced by bulk planetary, stellar, and/or orbital parameters: radius, mass, irradiation, etc.   However, the complex interplay of all these factors --- along with atmospheric models that do not yet encapsulate all necessary processes --- means that achieving this goal is an extremely  complicated task. 

For now we consider a simpler question: under what conditions do the atmospheres of warm Neptunes show detectable spectral features in transit? This is a lower-order question than those enumerated above, but detecting spectral features is a necessary first step toward these more ambitious goals. For this purpose our choice of WFC3/G141 observations is ideal, since  a single species --- \water --- dominates the expected opacity at these wavelengths. Although \methane, HCN, and other species  absorb in the G141 bandpass,  none of these have been reported in our sample. 

\begin{figure}
%%\plottwo{f2.eps}{f2_color.eps}
\includegraphics[width=4in]{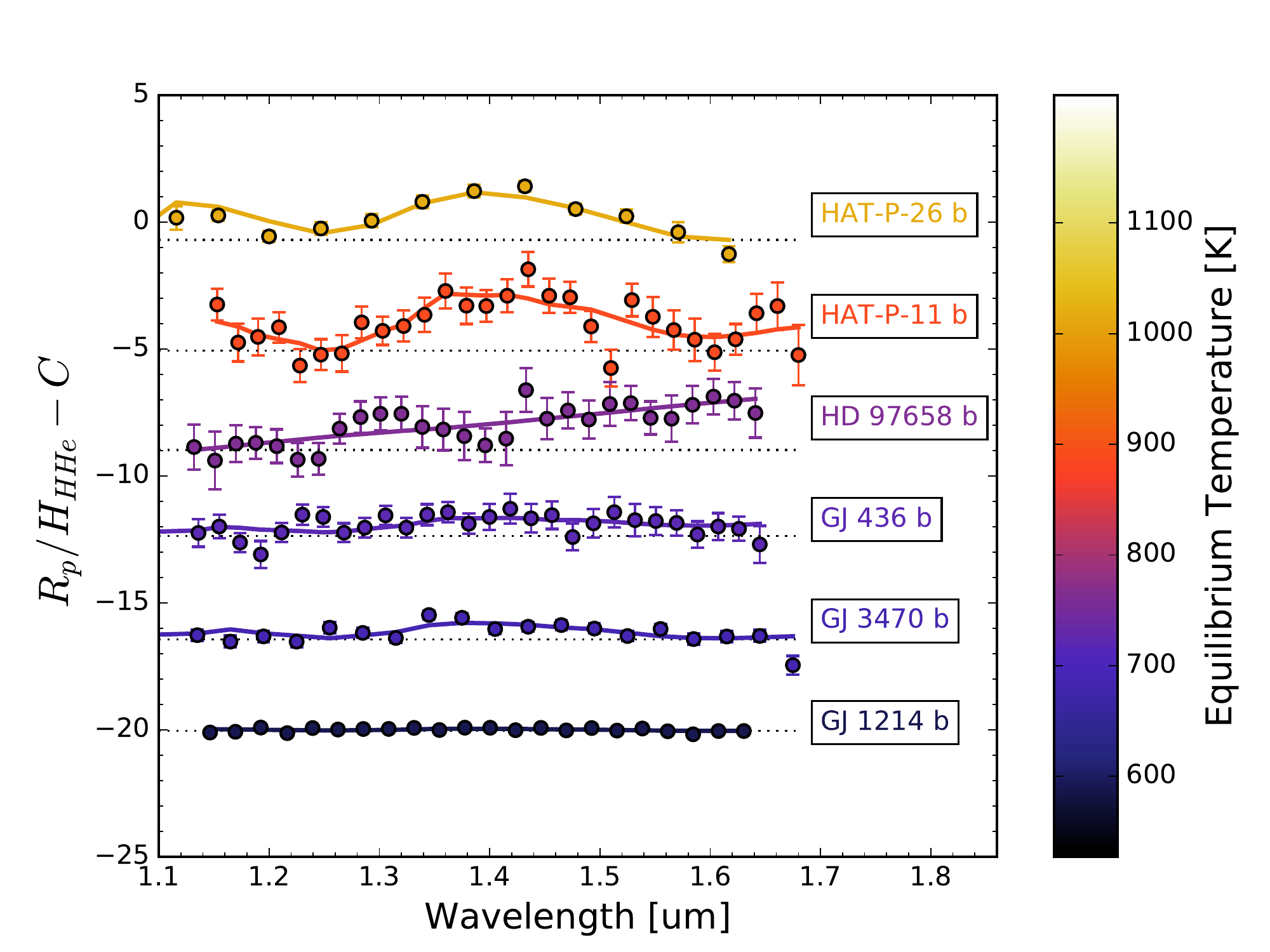}
\caption{Near-IR transmission spectra of the six warm Neptunes
  observed with HST/WFC3 (points) compared to illustrative models
  (lines), in units of scale height $H_{HHe}$ (assuming $\mu$\,=\,2.3
  g~mol$^{-1}$).  Data are from
  \cite[][]{fraine:2014,knutson:2014a,knutson:2014b,kreidberg:2014,wakeford:2017};
  some error bars are smaller than the plotted
  points. A cloud-free, Solar-composition atmosphere would produce a spectral amplitude of about 6.7 scale heights.\label{fig:spectra}}
\end{figure}

To facilitate comparison of the WFC3/G141 measurements of our diverse targets, we place all the spectra on the same scale. A fundamental unit in transmission spectroscopy is the atmospheric scale height $H= k_B T_{eq} / \mu g$, where $g$ is the planet's surface gravity. Here $\mu$ is the mean molecular weight, 2.3~amu for a H-dominated atmosphere with Solar abundances and  increasing only slowly for M/H up to $\sim100\times$ Solar (where $\mu=3.05$~amu); beyond this, $\mu$ rises increasingly quickly with increasing M/H.  We therefore initially assume $\mu=2.3$ for all our sample and calculate $H_{HHe}$, the scale height assuming a low atmospheric metallicity.  If any of our targets have highly enriched atmospheres, $H < H_{HHe}$ and the true signal amplitude would be a larger number of scale heights. We plot the normalized WFC3/G141 measurements  in Fig.~\ref{fig:spectra}. One conclusion is immediately apparent: not all small, cool exoplanets have flat spectra.

To estimate the amplitude of H2O absorption, we fit a template spectrum $S_t$ normalized to unit amplitude. Using weighted linear least squares, we fit the observed spectrum $S_o = a_1+ a_2\lambda + a_3 S_t$, where $\lambda$ is the wavelength in microns, and the $a_i$ are constants.  The template comes from a carbon-free atmospheric model of GJ~1214b, described by \cite{crossfield:2011}. When the template is convolved to a resolution of 600, approximately 6.7 scale heights separate the peak at 1.4\,$\mu$m from the trough at 1.25\,$\mu$m\footnote{This template spectrum is available for download as an electronic supplement to the paper.}. The resulting scaled best-fit models are plotted over the measurements in Fig.~\ref{fig:spectra}, and the
\water\ feature amplitudes are listed in Table~\ref{tab:targets}.

We then investigate whether the amplitude of \water\ absorption in these planets' spectra could be explained by various planetary and system parameters.  We investigated planetary $R_P$, $M_P$, $\rho_p$, $g$, $T_{eq}$, and bulk H/He mass fraction \citep[$f_{HHe}$; from ][]{lopez:2014},   stellar $T_{eff}$, and predicted FUV and XUV irradiation \citep[from ][]{france:2016}. We then simply investigate which of these quantities correlates with our measured amplitude for the \water\ feature, both by computing the Pearson correlation coefficient $r$ and associated chance probability $p$, and  by fitting a linear relation and comparing the resulting $\chi^2$. 

\begin{figure}
%%\plottwo{f2.eps}{f2_color.eps}
\includegraphics[width=4in]{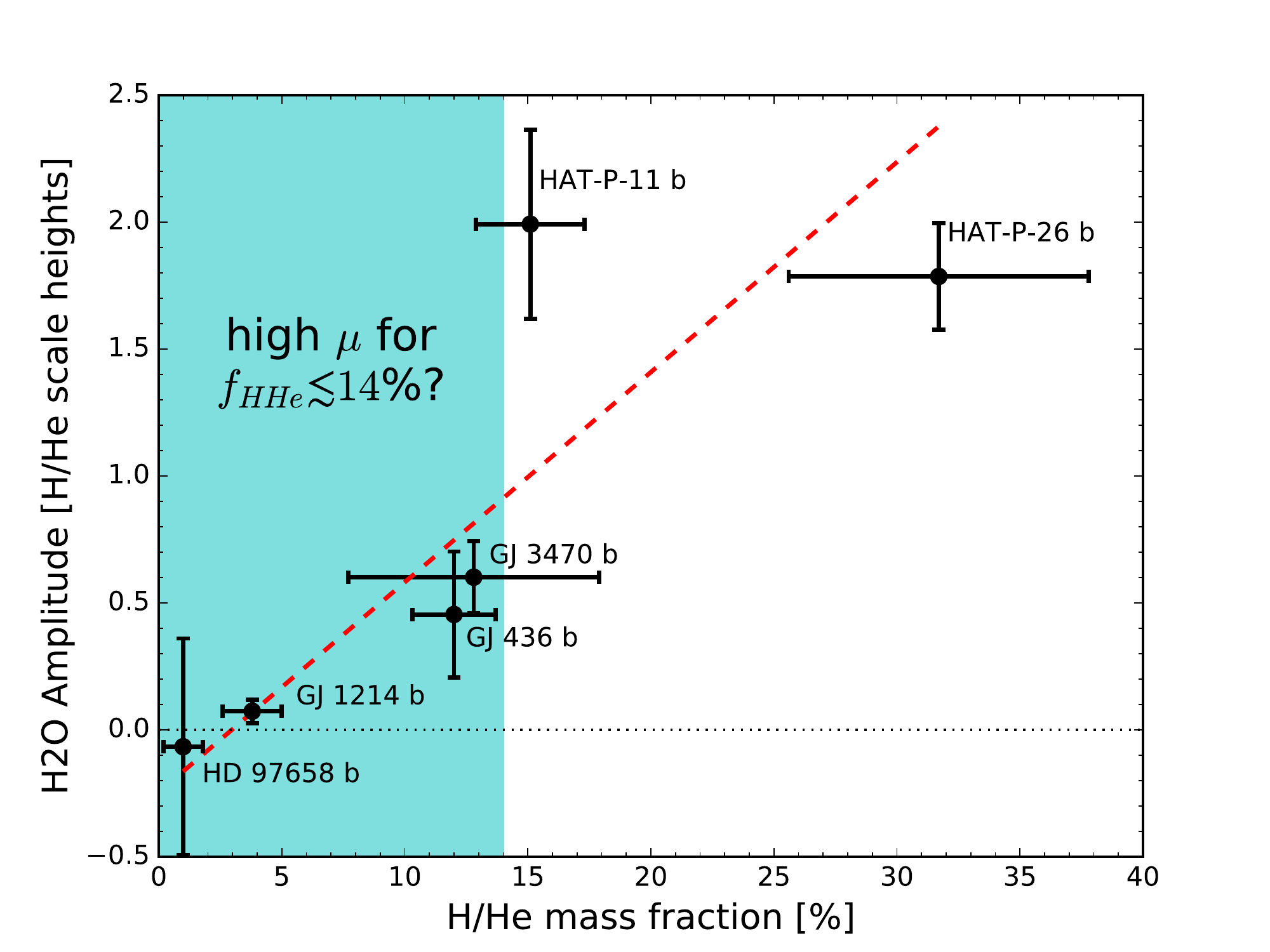}
\caption{Correlation of H/He mass fraction \citep[$f_{HHe}$, from][]{lopez:2014} with the measured amplitude of the WFC3/G141 \water\ features shown in Fig.~\ref{fig:spectra}, along with the best-fit linear trend (red dashed line).  This correlation suggests that smaller planets (with lower $f_{HHe}$) might have higher-metallicity atmospheres. \label{fig:hfrac}}
\end{figure}

\begin{figure}
%%\plottwo{f2.eps}{f2_color.eps}
\includegraphics[width=4in]{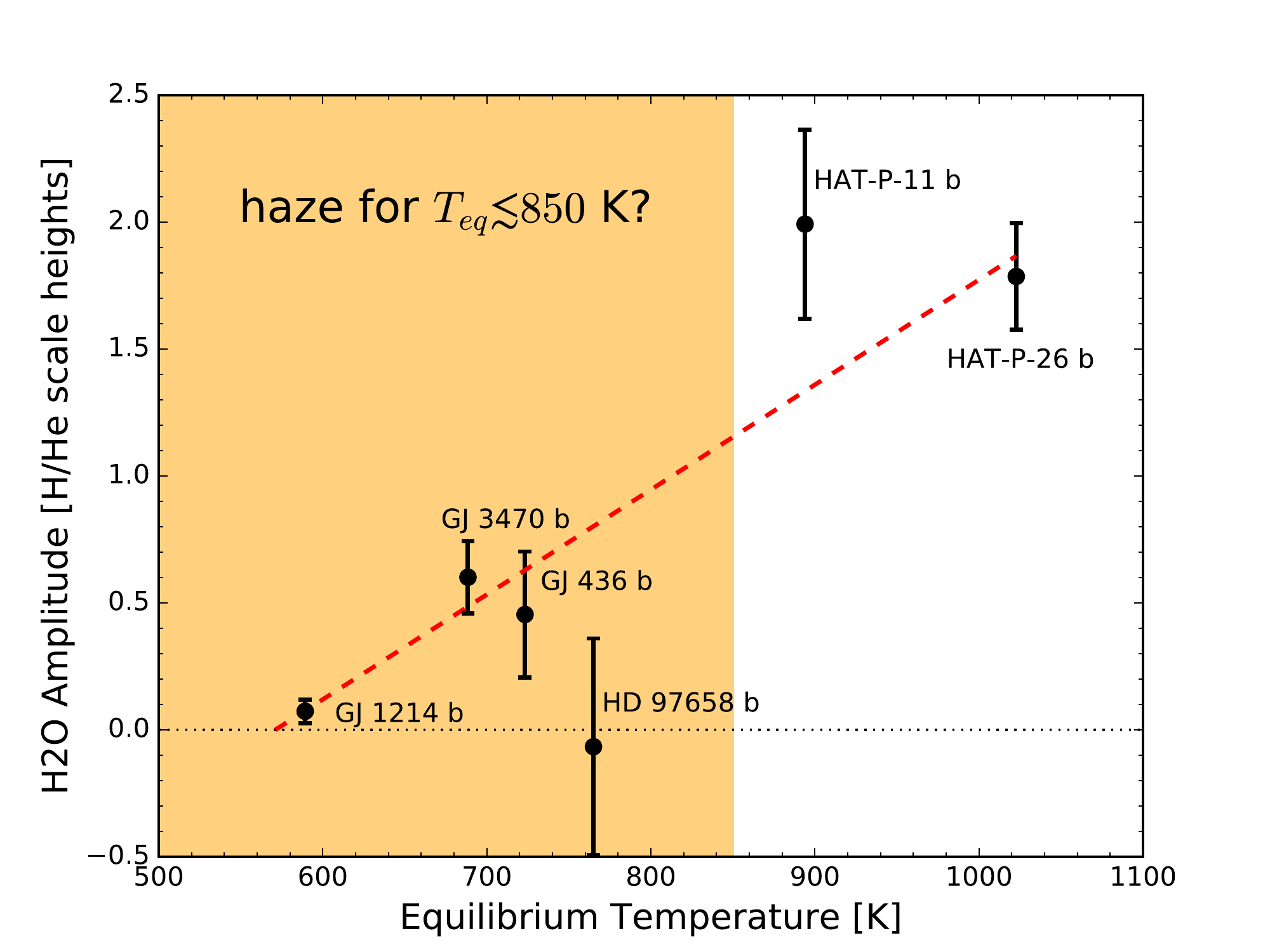}
\caption{Correlation of planetary equilibrium temperature with the measured amplitude of the WFC3/G141 \water\ features shown in Fig.~\ref{fig:spectra}, along with the best-fit linear trend (red dashed line). This correlation suggests that hazes might become more significant for planets with $T_{eq} \lesssim 850$~K. \label{fig:teq}}
\end{figure}

\section{Results}
\label{sec:results}
Table~\ref{tab:fits} summarizes how a variety system properties
correlate with the detectability of atmospheric features in warm
Neptunes. Two properties stand out as better predictors than the others. The first is $T_{eq}$, which gives the lowest
$\chi^2$ and the second-highest $r$ (with $p=4\%$). The second possibility
is the pair $R_P$ and $f_{HHe}$ (which both have $p<5\%$). We note that $R_P$ and $f_{HHe}$ are expected to be tightly correlated for planets of this type \citep{lopez:2014}. Since $f_{HHe}$ is more physically linked to atmospheric composition and is directly
connected to the observable atmosphere probed by transmission
spectroscopy, we henceforth consider only $f_{HHe}$ along with
$T_{eq}$.  The correlation plots for these two parameters are shown in
Fig.~\ref{fig:hfrac} and~\ref{fig:teq}, respectively, and the
implications for each are discussed below.

\begin{deluxetable}{l r r r}
\tablecaption{  Correlations \label{tab:fits}}
\tablehead{
\colhead{Parameter}  &  \colhead{$r$} & \colhead{$p$}   & \colhead{$\chi^2$} %\tablenotemark{a}
}
\startdata
$T_{eq}$      & 0.83 & 0.040 &  6.1 \\
$R_P$         & 0.86 & 0.027 &  8.0 \\
$f_{HHe}$     & 0.81 & 0.049 &  8.2 \\ \hline
$\rho_P$      &-0.69 & 0.13  &  7.1 \\ 
$M_P$         & 0.74 & 0.092 & 11.7 \\
$T_{eff}$     & 0.49 & 0.32  & 13.4 \\
$g_P$         &-0.44 & 0.38  & 10.8 \\
$\log$(FUV)   &-0.12 & 0.82  & 13.0 \\
$\log$(XUV)   &-0.30 & 0.56  & 13.3 \\
  \enddata
%  \tablenotetext{a}{Four degrees of freedom.}
\end{deluxetable}

\subsection{A correlation with H/He mass fraction}
The correlation between \water\ feature scale height and $f_{HHe}$ is
shown in Fig.~\ref{fig:hfrac}. The higher the hydrogen/helium mass
fraction, the larger the features tend to be.  One possible
interpretation for this trend is that atmospheres with smaller H/He
envelopes have higher metallicity (and thus the true scale height $H$
is less than our predicted scale height $H_{HHe}$). This interpretation agrees with predictions from planet
formation models that smaller envelopes are more polluted by infalling
planetesimals \citep{fortney:2013, venturini:2016}.

If the amplitude of features depends solely on atmospheric
metallicity, we can use the observed spectra to put a lower limit on
the atmospheres' mean molecular mass. We assume that the feature
amplitude is linearly proportional to $\mu$ and that HAT-P-11b and
HAT-P-26b, our two largest planets, have $\mu=2.3$~amu (though neither
planet's spectrum is consistent with a Solar-metallicity, cloud-free
atmosphere).  Under this assumption GJ~3470b, GJ~436b, and GJ~1214b
would have $\mu$ equal to $8\pm2$, $10^{+9}_{-4}$, and
$61^{+63}_{-24}$, respectively.  Excluding non-physical negative
values, HD~97658b would have $\mu>3.0$ at 99.7\% confidence, consistent with previous analyses \citep{knutson:2014b}.

The derived values for GJ~3470b and GJ~436b are plausible; however,
the transmission spectrum of GJ~1214b is so flat that it requires a
higher metallicity composition than is expected for any volatile
species \citep{kreidberg:2014,morley:2015}. Furthermore, we confirm
that the data for GJ~1214b statistically justify a perfectly flat
spectrum, rather than inclusion of a \water\ signature; indeed, our
simplistic \water-only analysis finds $\mu>22$ with 99.7\% confidence.
Therefore the only explanation for these data is a high-altitude
condensate blocking the transmission of stellar flux.  This result may
therefore point toward a correlation between transmission spectral
amplitude and planetary equilibrium temperature rather than $\mu$, as
described next.

%According to this simplistic analysis all our planets' 1--1.7\,\micron\ transmission spectra could be explained by high-$\mu$ atmospheres. This trend could be tested by measuring M/H and $\mu$ for the planets in our sample via transit and/or eclipse observations. Unfortunately, current data constrain these quantities only weakly.

\subsection{A correlation with equilibrium temperature}
The correlation between \water\ feature scale height and $T_{eq}$ is shown in Fig.~\ref{fig:teq}. Instead of invoking higher M/H for the flatter spectra, an increasing particulate density of cloud or haze particles at high altitude could be suppressing features in the spectra, as inferred for GJ~1214b \citep[e.g.,][]{kreidberg:2014}.
 
A plausible route for forming high-altitude aerosols is through photochemical interactions with hydrocarbons (e.g. CH$_4$, C$_2$H$_2$), producing hazes analogous to those seen on Titan.   \cite{morley:2015} modeled the interaction of stellar irradiation and atmospheric metallicity on the amount of high-altitude, high-order hydrocarbons, deemed ``soot precursors,'' and found a strong increase in the high-altitude abundances of these compounds as planetary equilibrium temperature drops over a narrow range from 1100~K to 800~K.  

The transition from haze-free to hazy atmospheres at temperatures of 800--1100~K, predicted by \cite{morley:2015}, matches the transition observed in Fig.~\ref{fig:teq} surprisingly well.  In this scenario,  the warmer HAT-P-11b and HAT-P-26b show strong features in transmission because their atmospheres are too warm to form obscuring photochemical hazes; the other warm Neptunes are below the critical irradiation threshold and so form sufficient haze to block most or all of the expected transmission signature.  This trend could be tested by observing additional warm Neptunes across this $T_{eq}$ range, and by more sophisticated modeling of haze formation in this atmospheric regime.

The correlation with $T_{eq}$ could instead indicate something about
cloud top pressure instead of total aerosol content, but the data do
not seem to bear this out.  When considering HST/WFC3 G141 data alone
and assuming a Solar-metallicity atmosphere, the top of an opaque
cloud deck in these planets' atmospheres lie at pressures of $\sim$0.1
mbar for GJ1214b \citep{kreidberg:2014}, $\sim$0.4 mbar for GJ~3470b
(Benneke et al., submitted), $\lesssim$1 mbar for HD~97658b
\citep{knutson:2014a}, $\sim$1 mbar for GJ~436b \citep{knutson:2014b},
and $\sim$100 mbar for HAT-P-11b \citep{fraine:2014}.  If all planets
had Solar-metallicity, then clouds would seem to sink deeper into the
atmosphere with increasing planet temperature --- the opposite of what
is typically understood to happen in brown dwarf atmospheres
\citep[e.g.,][]{lodders:2004}. If temperature is related to
cloudiness, the connection to photochemical haze production seems more
likely.

%\section{Selection Effects and Biases}
%HAT-P-26b was targeted for HST/WFC3 specifically because it is low density and expected to have high H2 content (D.~Sing et al, priv.\ comm.).
%
%There are no cool, puffy planets because (a) if they orbited roughly Sun-like stars their transit depths would be shallow and transit would occur infrequently, so unlikely to be detected; (b) if they orbited low-mass stars, ... well, such planets are intrinsically quite uncommon compared to relatively smaller planets.
%
\section{Atmospheric Characterization with JWST}
\label{sec:tess}
At present, both our identified trends rely on a small sample of only
six planets. This is because just a handful of Neptune-size and
smaller planets are feasible targets for atmosphere characterization
with current facilities.  JWST will revolutionize the study of planets
in this size regime, thanks to its larger aperture and broader
wavelength coverage, enabling higher signal-to-noise characterization
of more spectral features. In addition, the Transiting Exoplanet
Survey Satellite (TESS) is predicted to discover hundreds of small
planets orbiting bright, nearby stars, which are ideal candidates for
atmospheric study \citep{ricker:2014,sullivan:2015}. In this section,
we explore the feasibility of characterizing the atmospheres of the
TESS population of warm Neptunes with JWST, based on the possible
trends identified above.

\subsection{Exposure Time Calculator}
The amplitude of features in a transmission spectrum is traditionally calculated using the atmospheric scale height and is $\propto H R_P/R_*^2$ \citep{miller-ricci:2009}. However, for cooler planets with potentially higher metallicities, this relation does not strictly hold. Particularly below 1000~K and for M/H  $>100\,\times$ solar, the dominant molecular species vary,  qualitatively changing the shape and amplitude of absorption features \citep{moses:2013}.

Here we introduce an empirical scaling relation to estimate the
observing time needed to make a significant (5$\sigma$) detection of
features in a JWST transmission spectrum. This relation is based on
model planet spectra calculated over a grid of atmospheric
temperatures ($400 - 1600\,\mathrm{K}$) and metallicities ($Z = 1 -
1000\times$ solar), using the open-source radiative transfer code
ExoTransmit \citep{kempton:2016}.  Our nominal planet has a transit
depth of $1\%$ and surface gravity $g = 10\,\mathrm{m/s^2}$, and so is
roughly comparable to HAT-P-11b or HAT-P-26b. Our calculation includes
opacity from the major absorbing species expected for gaseous
planetary atmospheres: H$_2$O, CH$_4$, CO, CO$_2$, and NH$_3$, in
addition to collisionally-induced H$_2$ absorption.

We use the PandExo tool \citep{batalha:2017b} to simulate {\em JWST} observations for the NIRISS instrument in Single Object Slitless Spectroscopy (SOSS) mode. NIRISS/SOSS yields the highest information content for any single JWST instrument/disperser combination \citep{batalha:2017a}.  For the star, we use a PHOENIX stellar model with a temperature of 4000 K, surface gravity log~$g = 4.5$, and  $J = 10$~mag. We assume the data are photon-noise limited (i.e., there is no noise floor) but the assumed noise floor is not critical, since we find that most TESS targets will orbit stars with $J\gtrsim8$~mag.
% In reality - bright stars will probably reach some noise floors. And may saturate!

For each planet in our grid, we calculate how many total hours of observing time (including transit and an equal amount of baseline) are needed to detect features in the transmission spectrum at 5$\sigma$ confidence on average. We determine the detection significance by simulating a sample of 100 model spectra, binning them to a resolution of $\lambda/\Delta_\lambda \sim 100$, and calculating the reduced $\chi^2$ values for the spectra compared to a flat line.  Based on these results, we fit an analytic model to the observing time in hours, $t_{hr}$,  required to distinguish a spectrum from a featureless flat line at $5\sigma$ confidence:

\begin{equation}
t_{hr} = A^2 F_*^{-1/2}
\end{equation}
where $F_*$ is the relative stellar photon flux,  $10^{-0.4(J-10)}$, and 
\begin{equation}
A = [(1.3 - (T_{eq}-360)^{0.0035} + 0.25Z/T)) g R_*^2/R_p]^2    
\end{equation}
where $R_p$ is the planet radius ($\times10^7$ km), $R_s$ is the stellar radius ($\times10^8$ km), and $g$ is the planet's surface gravity (in 10~m~s$^{-2}$).

%We note that 
The functional form of the above relation is not physically motivated,
but it successfully reproduces the exposure time needed to an accuracy
of 15\% on average over the whole parameter space we consider
($T_{eq}=400$--1600~K, M/H\,=\,1--1000).

\begin{figure*}
%%\plottwo{f2.eps}{f2_color.eps}
\includegraphics[width=2.5in]{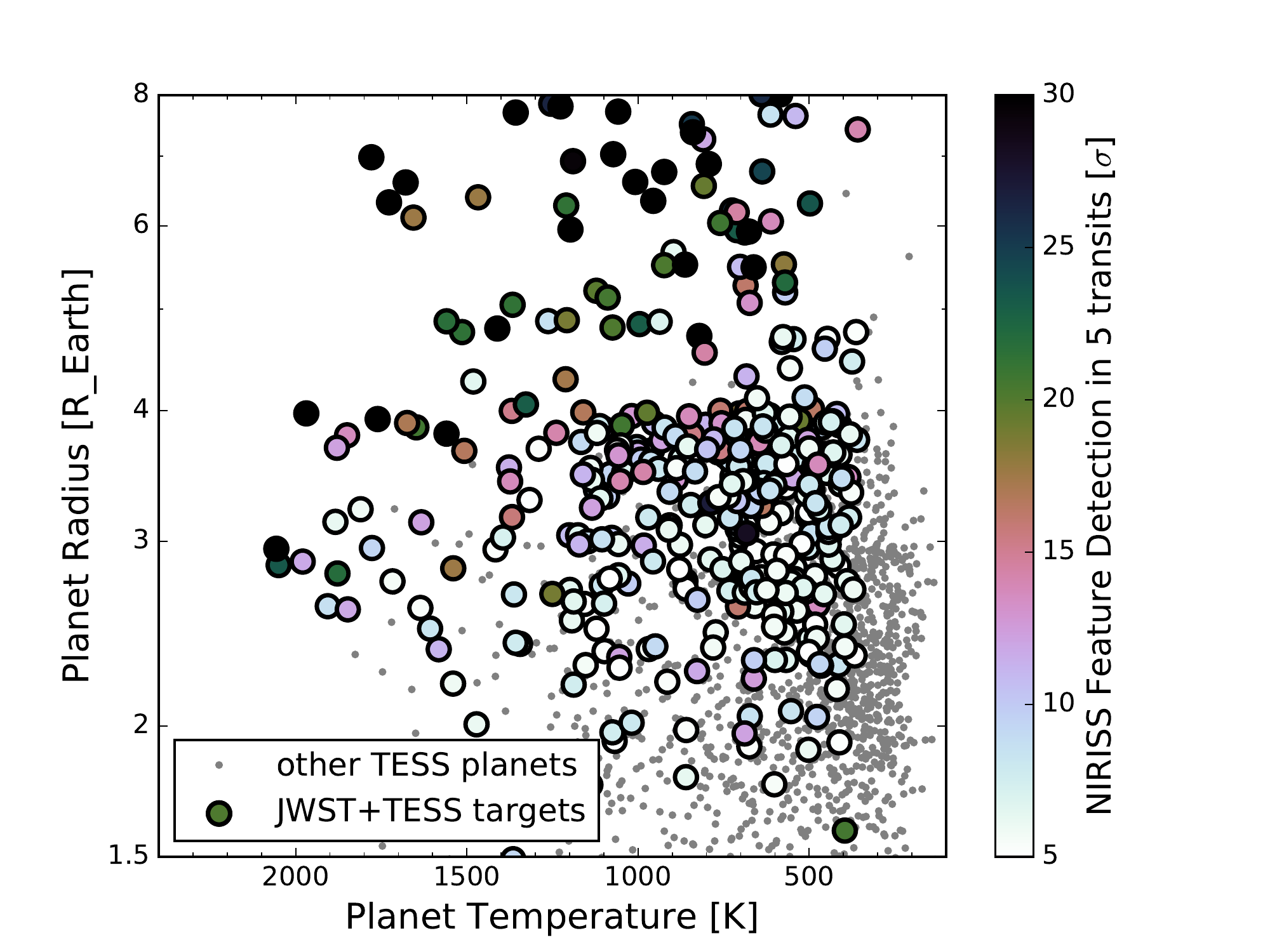}
\includegraphics[width=2.5in]{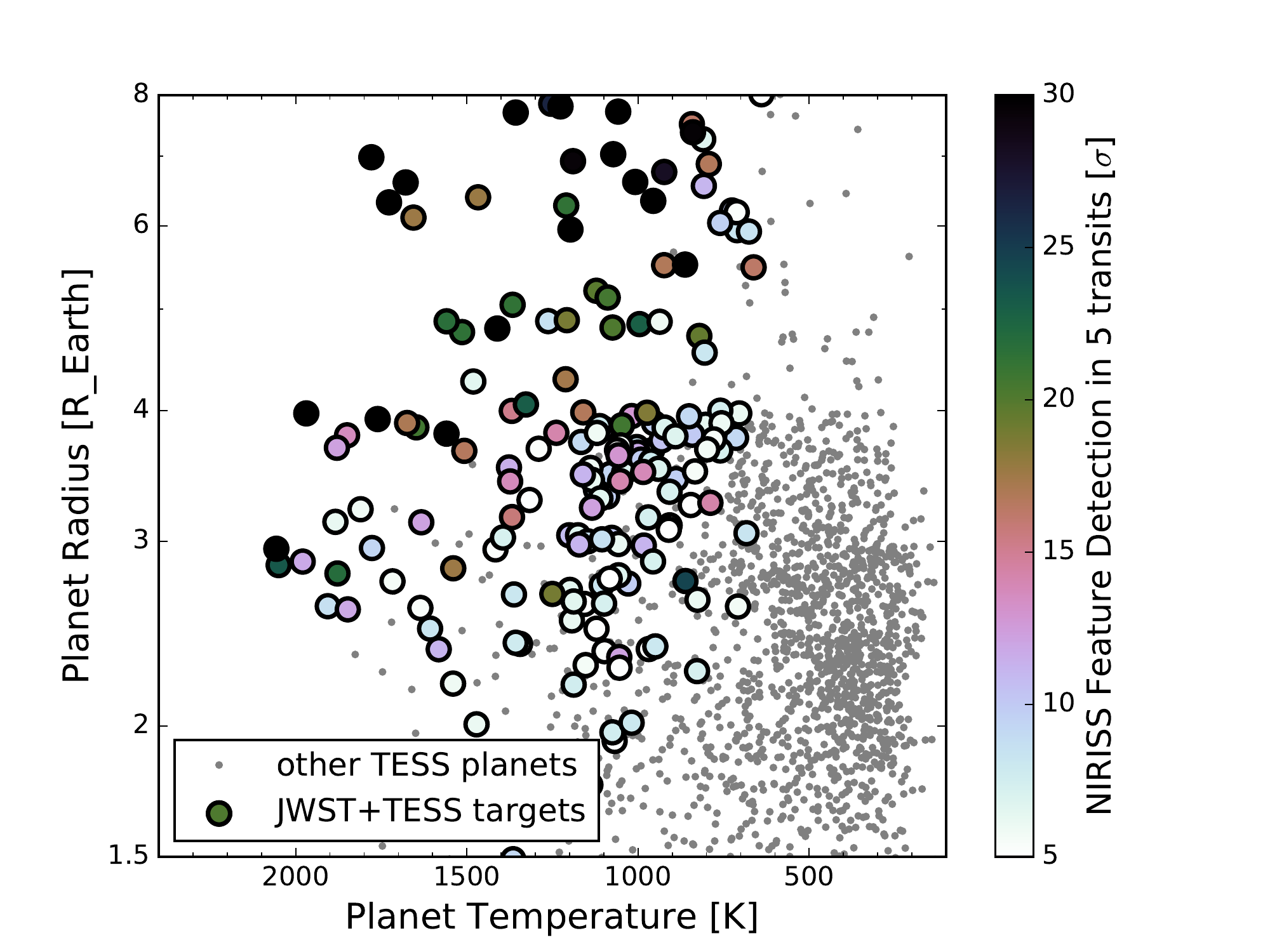}
\includegraphics[width=2.5in]{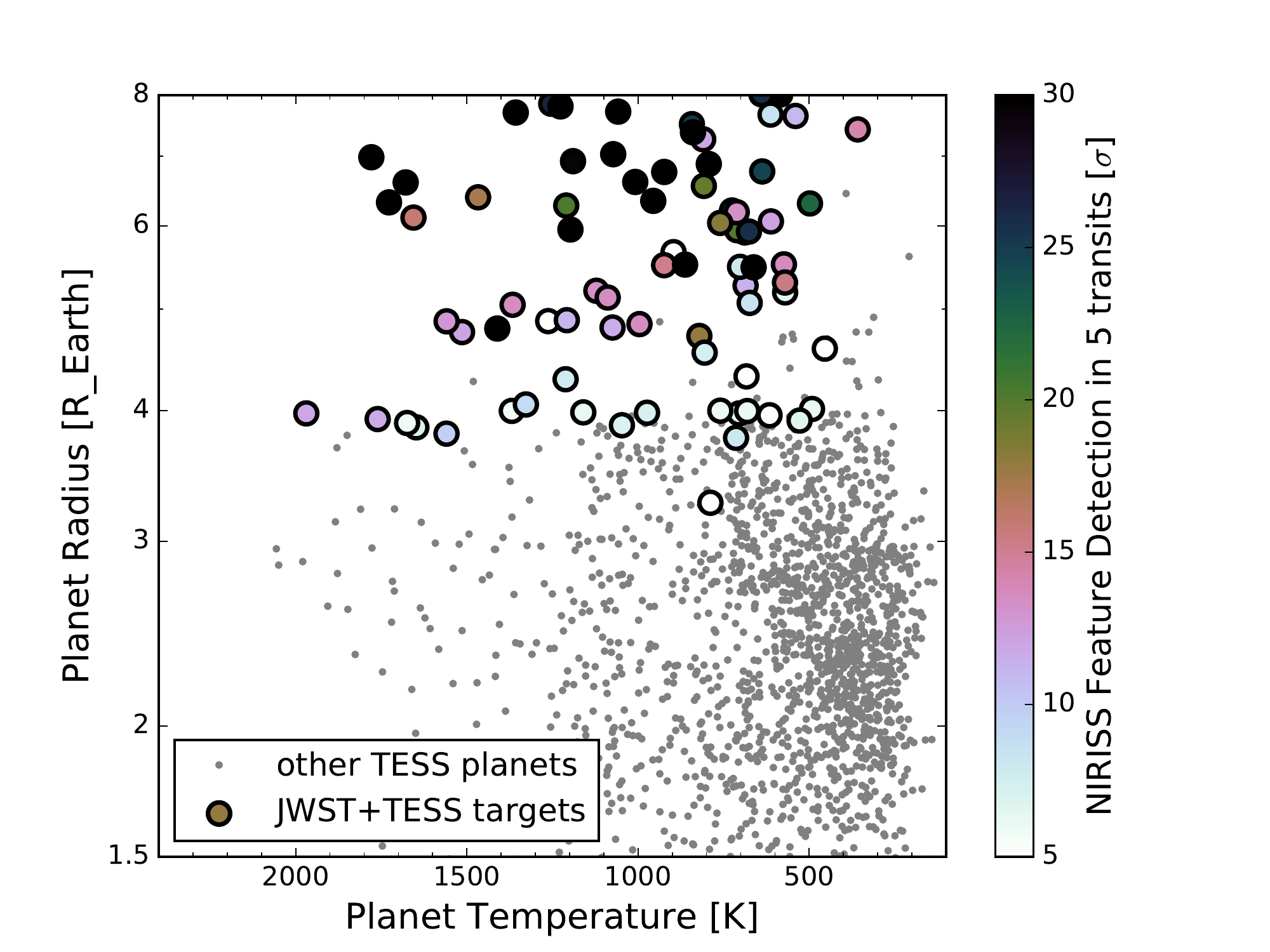}
\caption{Expected TESS planets accessible to JWST/NIRISS transmission spectroscopy. Large points are those TESS planets for which NIRISS  could distinguish spectral features from a flat line at $5\sigma$ in $\le5$ transits. Small points show the rest of the expected TESS sample \citep[from][]{sullivan:2015}. {\em Left:} The nominal case as modeled by ExoTransmit; the indicated JWST+TESS sample comprises 370 planets smaller than $6R_\oplus$. {\em Center:} Assuming that the amplitudes of transmission spectra  decrease linearly with $T_{eq}$ as shown in Fig.~\ref{fig:teq}; in this case, the sample drops to 154 planets. {\em Right:} Assuming that transmission amplitude decreases linearly with $R_P$ (a proxy for $f_{HHe}$) as shown in Fig.~\ref{fig:hfrac}; in this case, the sample drops to just 47 planets smaller than $6R_\oplus$.
\label{fig:tess_nominal}}
\end{figure*}

%
%\subsubsection{Added IJMC, 2017/07/13:}
\subsection{TESS Yield of JWST-accessible Neptunes}
We use this relation to estimate the number of expected {\em TESS}
planets whose transmission spectra {\em JWST}/NIRISS can distinguish
from a flat line at $\ge 5 \sigma$ in five transits.  The simulated
TESS yield of \cite{sullivan:2015} does not include an assumption for
each planet's atmospheric M/H, so we initially make the optimistic
(indeed, unrealistic) assumption that M/H\,=\,3 for all planets. We
then subsequently assume that the transmission spectral amplitudes of
these planets scales linearly with $T_{eq}$ or $R_P$ (a proxy for
$f_{HHe}$) as shown in Figs.~\ref{fig:hfrac} and~\ref{fig:teq}.
Fig.~\ref{fig:tess_nominal} shows the result of this investigation
under these three assumptions.

In the nominal case the expected accessible sample comprises 370
planets with $R_P<6R_\oplus$, suggesting a large haul of interesting
planet targets. However, even under this most favorable case the
number of systems amenable to very high-S/N atmospheric measurements
is much smaller: it drops to just 30 planets when the bar is raised to
20$\sigma$. The $5\sigma$ {\em TESS}+{\em JWST} sample also decreases
rapidly once we account for the scalings implied by our analyses. The
sample drops by over half, to 154 planets, when the signal decreases
linearly with $T_{eq}$: from full amplitude at 1000~K to featureless
at 550~K. If instead transmission amplitude decreases with $R_P$ from
$6R_\oplus$ to $2.5R_\oplus$ (using $R_P$ as a proxy for $f_{HHe}$),
the sample shrinks to just 47 planets. Under this pessimistic
assumption, almost no planets with $R_P\gtrsim4 R_\oplus$ are
accessible even after observing five transits with JWST/NIRISS; this
is because (by construction) the amplitude of transmission features
rapidly decreases for these smaller planets.

In all of these simulations TESS identifies warm Neptunes with a range
of properties, from those observable in just a few JWST transits to
those requiring many, many transits to plausibly detect any
atmospheric signatures (see Fig.~\ref{fig:tess_nominal}).  Until we
understand the atmospheric processes dominating these planets, the
best targets for detecting features in transmission spectroscopy will
be warmer planets with larger $f_{HHe}$ and $R_P$.

\section{Conclusions}
\label{sec:conclusion}
We have identified two possible trends in the transmission spectra of
warm Neptunes, both at $>95\%$ confidence. As shown in
Figs.~\ref{fig:hfrac} and~\ref{fig:teq}, the amplitude of
\water\ absorption at 1.4\,\micron\ increases with both $T_{eq}$ and
$f_{HHe}$ (or equivalently, $R_P$) for the 2--6\,$R_\oplus$,
500--1000~K planets in Table~\ref{tab:targets}.  The scaling with
$f_{HHe}$ could indicate that smaller planets have higher-metallicity
atmospheres, but at least for GJ~1214b we know that increased M/H
cannot explain the observations: aerosols must be involved
\citep{kreidberg:2014}. Presumably both metallicity and aerosol
production both play a role, though the relative contribution of each
effect remains undetermined.

%We therefore favor the trend with
%$T_{eq}$, which is consistent with models predicting increasing haze
%formation below $\sim$900~K \citep{morley:2015}.

The trends we have identified remain tentative due to our small sample
size. The two key parameters of $T_{eq}$ and $f_{HHe}$ are largely
degenerate in our half-dozen planets. Aside from the more poorly
characterized HD~97658b, our coolest planet (GJ~1214b) is also the
smallest, and the largest and most H/He-rich planet (HAT-P-26b) is
also the hottest.  Obtaining transmission spectra of cool yet puffy
Neptunes \citep[such as
  HD~3167c;][]{vanderburg:2016b,christiansen:2017} and/or hotter yet
lower-$f_{HHe}$ planets \citep[such as
  HIP~41378b;][]{vanderburg:2016c} will be essential if we are to
break this degeneracy and truly determine the key factors controlling
these planets' atmospheres. Additional measurements for planets with
less precise transmission spectra (such as HD~97658b) would also help
with this effort.

We also estimated the number and type of warm Neptunes to be
discovered by {\em TESS} that will also be accessible to {\em JWST}
transmission spectroscopy. Fig.~\ref{fig:tess_nominal} shows that
under the common, optimistic assumption of cloud-free conditions, {\em
  JWST}/NIRISS could distinguish atmospheric features in $\sim$370
warm Neptunes by observing $\le$5 transits of each.  But experience
shows that the true accessible yield will be lower: just $\sim$150 if
feature amplitude scales with $T_{eq}$, and only $\sim$50 if it scales
with $R_P$ or $f_{HHe}$. Until the physical mechanism(s) underlying
these trends can be identified, the most promising warm Neptunes for
transmission spectroscopy are those with $R_P\gtrsim 4 R_\oplus$ and
$T_{eq}\gtrsim850$~K.  Of course, such observations cannot alone
reveal the nature of the trends discussed here; for that, a larger and
more diverse planetary population must be explored.

Until {\em TESS} and {\em JWST} arrive, more progress can be made by observing additional warm Neptunes in transit, both with {\em HST} and with ground-based spectroscopy.  Transit surveys continue to identify new targets, and scheduled or pending observations of planets such as WASP-107 (GO-14915, PI Kreidberg), Kepler-51 (GO-14218, PI Berta-Thompson), K2's warm Neptunes (GO-15333, PIs Crossfield \& Kreidberg), and others will help reveal the trends in atmospheric properties of Neptune-size exoplanets.

%\textbf{We propose to significantly expand  the sample of mini-Neptunes with atmospheric characterization by observing four  recently discovered  exoplanets all in multi-planet systems.} Our low-mass targets  (see Fig.~\ref{fig:Target-planets}) range from 5--20~$M_{\oplus}$ and have equilibrium temperatures ($T_{eq}$) of $600-1000$~K. Their bright host stars give \textbf{these planets  the highest S/N for atmospheric study of any mini-Neptunes yet to be explored.} We will acquire precise spectra that will determine the metallicities, C/O ratios, and cloud properties of all these planets. With these observations we will test models of planet mass-metallicity correlations, planet formation and migration, and cloud physics.  Our study will elucidate the transitional regime between rocky planets to  gas giants, prepare for future JWST observations, and prioritize  follow-up of  similar systems discovered by TESS.

\acknowledgements We thank Eliza Kempton, Ruth Murray-Clay, and David
Sing for productive discussions and encouragement.  We also thank the
organizers of the 2017 Disks and Planets conference at Ringberg for
motivating us to consider these issues in more depth, and our referee
for careful attention that improved the quality of this work.

% Harvard Society of Fellows

\bibliographystyle{apj}
%\bibliography{../ms}

%\begin{figure}
%\includegraphics[width=5in]{example_spectrum.pdf}
%\caption{An example simulated spectrum for a 600 K, $100\times$ solar metallicity planet %\label{fig:jwst_spectrum}. The uncertainties are based on a PandExo simulation for the NIRISS instrument. }
%\end{figure}

%\clearpage

\end{document}